\documentclass{article}
\usepackage{amsmath}
\usepackage{graphicx}
\usepackage{authblk}
\numberwithin{equation}{section}
\newcommand{\keywords}[1]{\textbf{\textit{Index terms---}} #1}

\begin{document}

\title{Sensitivity for detection of decay of dark matter particle using ICAL at INO}
\author[1,2]{ N. Dash\thanks{nitali.dash@gmail.com}}
\author[1,2]{V. M. Datar\thanks{vivek.datar@gmail.com}}
\author[3]{ G. Majumder\thanks{majumder.gobinda@gmail.com}}

\affil[1]{Nuclear Physics Division, Bhabha Atomic Research Centre, Mumbai - 400085, INDIA}
\affil[2]{Homi Bhabha National Institute, Anushaktinagar, Mumbai - 400094, INDIA}
\affil[3]{Tata Institute of Fundamental Research, Mumbai - 400005, INDIA}

\maketitle

\begin{abstract}
We report on the simulation studies on the possibility of dark matter particle (DMP) decaying into leptonic modes. While not much is known about the properties of dark matter particles except through their gravitational effect, it has been recently conjectured that the so called ``anomalous Kolar Events" observed some decades ago may be due to the decay of unstable dark matter particles (M.V.N. Murthy and G.Rajasekaran, Pramana, {\bf 82}, 609 (2014)). The aim of this study is to see if this conjecture can be verified at the proposed Iron Calorimeter (ICAL) detector at INO. We study the possible decay to leptonic modes which may be seen in this detector with some modifications. For the purposes of simulation we assume that each channel saturates the decay width for the mass ranging from $1-50~\rm{GeV/c^2}$. The aim is not only to investigate the decay signatures, but also, more generally, to establish lower bounds on the life time of DMP even if no such decay takes place.
\end{abstract}


\section{Introduction}
It is now established that the dark matter particles (DMPs) constitute $80-85\%$ of all matter in the Universe, but we know very little about the properties of these particles. Even though they are non-luminous in nature, their presence has been inferred through their gravitational effect. The presence of Dark matter was investigated in detail by Vera Rubin using their kinematical study of the galaxies \cite{rubin} during $1960-1970$ using the rotational curves. Further evidence came recently through the study of gravitation lensing. It is known that most of the DMPs must be non-baryonic and they do not have electromagnetic interaction.

They may be composed of hypothetical neutral particles such as Weakly Interacting Massive Particles (WIMPs), axions, or sterile neutrinos which are beyond the purview of the Standard Model. While properties such as spin and parity of DMPs are completely unknown, there are theoretical estimates of their interaction cross-section with nuclei and their life time. Cryogenic bolometer based experiments like CDMS \cite{cdms}, LUX \cite{lux}, CRESST \cite{cresst} are trying to detect it directly through their recoil against the detector nuclei. The indirect space based experiments like PAMELA \cite{pamela}, Fermi-LAT \cite{fermilat}, AMS \cite{ams} are trying to detect the DMP by measuring the decay or annihilation products in the form of excess anti-matter in the Universe. Similarly the water and ice cherenkov based detectors like Super-Kamiokande \cite{superk} and IceCube \cite{icecube} have put limit on the life time of the DMP by detecting the neutrino signal, mostly from dark matter annihilation in the centre of the Sun, the centre of the Earth and in the Galactic centre. Both direct and indirect experiments have provided some hints of DMPs.

It is not known whether they are stable or not. However, if they are unstable their life time should be of the order of or longer than the age of the Universe. It has been recently conjectured that some anomalous events observed at the deep underground laboratory at Kolar Gold Fields (KGF) in south India, the so called {\it Kolar events}, may be due to the decay of unstable dark matter particles whose mass is in the range of several $\rm{GeV/c^2}$ \cite{murthy}. These events, discussed in Refs.\cite{pramana,nimb,berkeley} have neither been confirmed nor shown to be spurious. They constitute $25\%$ of the total recorded events over two decades at a depth of 2.3~km. Immediately after the observation of these anomalous events, they were interpreted as the decays of a hypothetical heavy particle \cite{heavy1,heavy2,heavy3} with life time of the order of $10^{-9}~\rm{s}$ and mass of around $2-5~\rm{GeV/c^2}$, possibly produced in the interaction of neutrino or anti-neutrino with the rock surrounding the detector. However, no evidence for such a particle was found in other experiments including those using neutrino beams \cite{cern,fermilab}. 

While these anomalous events have evaded any conventional, standard model based, interpretation until now, the conjecture that the anomalous events are due to the decay of unstable dark matter opens up a new possibility. This may be established by future neutrino detectors, like the Iron Calorimeter (ICAL) at India-based Neutrino Observatory (INO) \cite{ino}. The sensitivity to dark matter decays may be enhanced by placing detector elements on the walls and ceiling of the large cavern housing ICAL.

In this paper we report on the simulation of the dark matter particle decaying into leptonic modes. For the purposes of simulation we assume that each channel saturates the width with the mass ranging from $1-50~\rm{GeV/c^2}$. The aim is not only to investigate the conjecture, but also, more generally, to determine the lower limits on the life time of DMP by assuming their number to be with in a limited volume of the ICAL cavern.

The ICAL detector proposed to be placed in a cavern at INO which is much larger than that at KGF. It is planned to construct a $50~\rm{kton}$ magnetised Iron Calorimeter under a rock cover of at least $1~\rm{km}$ all around. The main goal of the ICAL is to study atmospheric neutrinos ($\nu_{\mu}$), in particular the mass hierarchy of neutrinos. However it can also be used for other purposes such as searching for magnetic monopole \cite{mmical}, to observe or set limits on the life time of DMP by considering their number in the ICAL cavern and also from the annihilation or decay products of the DMP by looking at the centre of the solar system. 

The manuscript is organised as follows: in Sec.$2$, we briefly discuss the methods used to detect DMP. In Sec.$3$, we focus on the proposed detector for DMP detection at ICAL cavern. In Sec.$4$ and $5$, we present the analysis and results of the simulation study for DMP decay in the ICAL cavern. We conclude the paper in Sec.$6$.


\section{Detection Mechanism of dark matter particle}

The search for DMP has been carried out both directly and indirectly. In the latter category one looks for the missing mass in events in particle colliders measured in a $4\pi$ detector assuming that can then be ascribed to the production of DMP. In the former category the search involves the detection of nuclear recoils resulting from DMPs colliding with the atoms in a detector. These detectors consist of rare gases such as argon or xenon in the liquid or gas phase, scintillators (NaI(Tl), $\rm{CaF_2}$ etc.), semiconductors (HPGe or Si) or cryogenic bolometers. In space based experiments one looks for the products of the annihilation or decay of the DMP. The experiments looking for high energy neutrinos, try to observe distinctive high energy neutrinos obtained from the annihilation or decay of the DMPs at the centre of the Sun, the Earth and the Galactic halo. Because the large number of the dark matter particle is involved there. Here we introduce yet another method to look for the possible decays of DMP. Since the DMP is present everywhere, a deep underground neutrino detector should be able to detect the possible decay of DMP provided the mass of DMP is in a suitable range and its decay products are easily detectable. If no such decay is observed, one should be able to put lower bounds on the possible partial life times after sufficiently long exposure. As pointed out in Ref.\cite{murthy}, based on the analysis of Kolar events, a large cavern located deep underground as in the case of neutrino detectors should be able to identify the DMP decay if the life time is around the age of the Universe. Either way this provides a novel way of putting limits on the DMP decay or its detection which has not been compared before. 

We consider the last scenario in this paper to detect DMP using the ICAL detector at INO with the assumption that it decays to standard model particles. If not we obtain appropriate limits. For convenience, in this simulation, it is assumed that DMP is a neutral scalar particle ($\Phi_{\rm{DM}}$) that decays to lepton pairs only and is guided by the results from different satellite based experiments. They have a significant excess in positron to electron ratio above $10~\rm{GeV/c^2}$ but not in the anti-proton to proton ratio. This mode of decay is also most suitable method for detecting DMP using the ICAL detector. The ICAL is a sampling calorimeter and is especially suitable for tracking muons which may arise from DMP decays to $\mu^+\mu^-$ pairs. Therefore, in general, we look for the decay modes of the type
\begin{equation}
\Phi_{\rm{DM}} \rightarrow {\ell^+ + \ell^-}~{(\ell = \rm{e}, \mu, \tau)}.
\end{equation}

In order to detect such events at INO, a detector configuration with ICAL as the central detector and some additional detectors around it is proposed. This is described in the next section.


\section{A Detector for Dark Matter Particle Decay} 

Dark matter is believed to be present everywhere. The biggest cavern at INO is the ICAL cavern having dimensions of $\sim$ $132~\rm{m}~\times~26~\rm{m}~\times~32~\rm{m}$ and a cavern volume $\sim$ $10^{11}~\rm{cm^3}$. As mentioned in Ref.\cite{murthy}, this large volume will lead to an increase in the number of detected DMP decay events which is around $\sim$ $1/\rm{Yr}$ based on the analysis of Kolar events. So to detect all the decay products in the form of visible particles, in simulation we have placed detectors on the 4 walls of the ICAL cavern.

The ICAL is a magnetized calorimeter consisting of $150$ layers of iron plates as the absorber each having a thickness of $5.6~\rm{cm}$ interleaved with an air gap of $4~\rm{cm}$ to accommodate the active detector element viz. the resistive plate chamber (RPC). The RPC uses a gas mixture of Freon (95.15$\%$), Iso-Butane (4.51$\%$) and $\rm{SF_6}$ (0.34$\%$). The gas mixture and the appropriate high voltage allow RPC to operate it in avalanche mode. The passage of a charged particle through the detector ionizes the gas medium which induces a signal and is collected by honey-comb patterned pick-up panels, which are placed orthogonally on either side of the detector. They gives X, Y co-ordinates of a hit point with a position resolution of $3~\rm{cm}$ and time resolution of $1~\rm{ns}$. The iron plates are magnetized with an average field of $1.3~\rm{Tesla}$ allowing a measurement of the curvature of the muon track. The proposed ICAL detector will occupy $52~\rm{m}~\times~16~\rm{m}~\times~15~\rm{m}$ of the ICAL cavern. In principle, the remaining space of the ICAL cavern can be used for additional DMP detector installations. One such simple configuration is proposed in this paper.

\begin{figure}[h]
\centering
 {\includegraphics[width=0.49\textwidth]{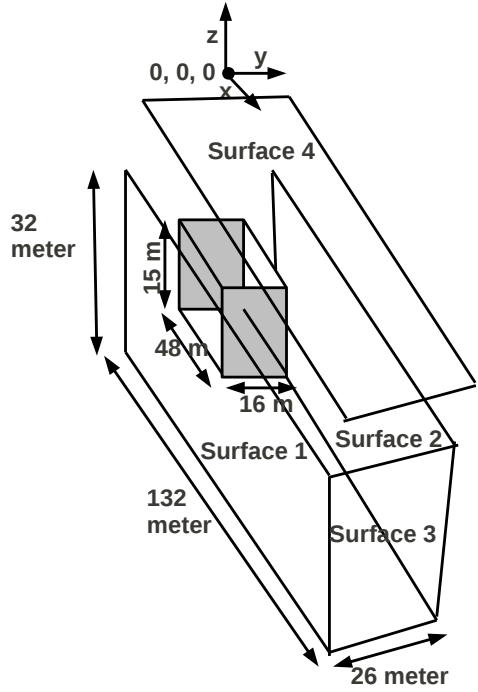}}
\caption{Schematic view of DMP detector with side detectors and 
the Iron Calorimeter.}
\label{dmpdetector}
\end{figure}

As the ICAL detector will be located towards one end of the cavern, four scintillator detectors (SDs) are mounted close to the walls of the cavern. A schematic diagram of the detector is as shown in Fig.{~\ref{dmpdetector}}. As shown in the figure surfaces 1, 2, 3 and 4 depict the scintillator detectors. Two of these detectors are placed along the length of the cavern with dimension of $132~\rm{m}~\times~0.04~\rm{m}~\times~32~\rm{m}$. The third one is placed in YZ plane with area of cross-section $26~\rm{m}~\times~32~\rm{m}$, having thickness of $4~\rm{cm}$. The fourth one is above the ICAL surface at a height of $17~\rm{m}$ from the top surface of the ICAL. This can also be used as a muon veto for ICAL. In the simulation only one layer of detector is used for each surface. So no energy measurement is available from these detectors. On the other hand they provide the signals for identifying back-to-back leptonic decays of DMPs. 

ICAL uses around $\rm{30,000}$ RPCs of dimensions $2~\rm{m}~\times~2~\rm{m}$. If RPC with same dimension were to be used instead of scintillator detector, then around $\rm{3,000}$ of them would be needed for a single layer lining on each plane. This is 10 times less than those required by ICAL.


\section{Simulation of decay of Dark Matter Particle}

The simulation is carried out in two regions separately. These two regions are specified by the air region i.e. the gap between the ICAL and the scintillator planes and another one is inside the ICAL detector. As mentioned in Sec.2, three decay channels of DMP are considered one at a time, with $\rm{100}\%$ branching ratio ($\rm{B}$) for each channel. These are
\begin{equation} 
\Phi_{DM} \rightarrow~{\rm{e}^+ + \rm{e}^-}~{(\rm{B} = 1)},
\end{equation}
\begin{equation}
\Phi_{DM} \rightarrow~{\mu^+ + \mu^-}~{(\rm{B} = 1)},
\end{equation}
and
\begin{equation}
\Phi_{DM} \rightarrow {\tau^+ + \tau^-}~{(\rm{B} = 1)}.
\end{equation}

If a DMP decays to $\mu^+\mu^-$ channel, it can be identified unambiguously in the ICAL provided its energy is 0.5 GeV or more, since there are two muon tracks back to back, which can be easily distinguished from other events. But for the other two channels i.e. $e$ and $\tau$, some uncertainties arise. As $\tau$ is the heaviest lepton with a short life time ($\sim~0.29\times10^{-12}~\rm{sec}$), its decay products include hadrons, muon, electron and their corresponding neutrinos. As the ICAL detector will use $56~\rm{mm}$ thick iron plates it will be difficult to separate electrons from $\pi$ decays. Due to the non-relativistic speed of DMP which is almost at rest, its decay will be isotropic. For a given mass of the DMP the energies of the daughter particles are obtained by two body kinematics. Assuming the mass of the decay particle and anti-particle pair to be m, the mass of the DMP to be M the momentum of each of the daughter particles, $\rm{p_1}$ and $\rm{p_2}$, is given by
\begin{equation} 
\label{mom} 
\vert\overrightarrow{\rm{p_1}}\vert = 
\vert\overrightarrow{\rm{p_2}}\vert = 
\sqrt{\frac{\rm{M}^2}{4} - 
\rm{m}^2}. 
\end{equation} 
Hence the DMP mass is taken as input to the simulation instead of the daughter particle energy. The mass of the DMP, decaying into $\mu$ and $e$ pairs, is varied from $1~\rm{GeV/c^2}$ to $50~\rm{GeV/c^2}$ in $1~\rm{GeV/c^2}$ steps. As the $\tau$ mass is $1.7~\rm{GeV/c^2}$, the DMP mass is varied from $6~\rm{GeV/c^2}$ to $50~\rm{GeV/c^2}$ in this decay channel. In the simulation two daughter particles start from a single vertex in opposite directions and with momenta given in Eq.(\ref{mom}). To obtain isotropic flux during the decay process the zenith angle ($cos\theta$) is smeared from 0 to $\pi$ and the azimuthal angle ($\phi$) by $2\pi$. The charge of one of the daughter particles in a decay event is chosen randomly and the other daughter particle has the opposite charge.


\subsection{Simulation in air region}

The High Energy Physics (HEP) simulation tool-kit GEANT4 \cite{geant4}, is used to do the simulation for DMP decay in the air region. In the simulation, the defined detector geometry is same as mentioned in Sec.2. The events are generated in the air gap i.e. between the side detectors and the ICAL detector. In a fraction of the simulated events the trajectories of the daughter particles are such that at least one of them enters the ICAL detector and its partner hits the side detector. Thus it will be possible to measure the energy of at least one decay product.

\begin{figure}[h]
\centering
\includegraphics[width=0.49\textwidth]{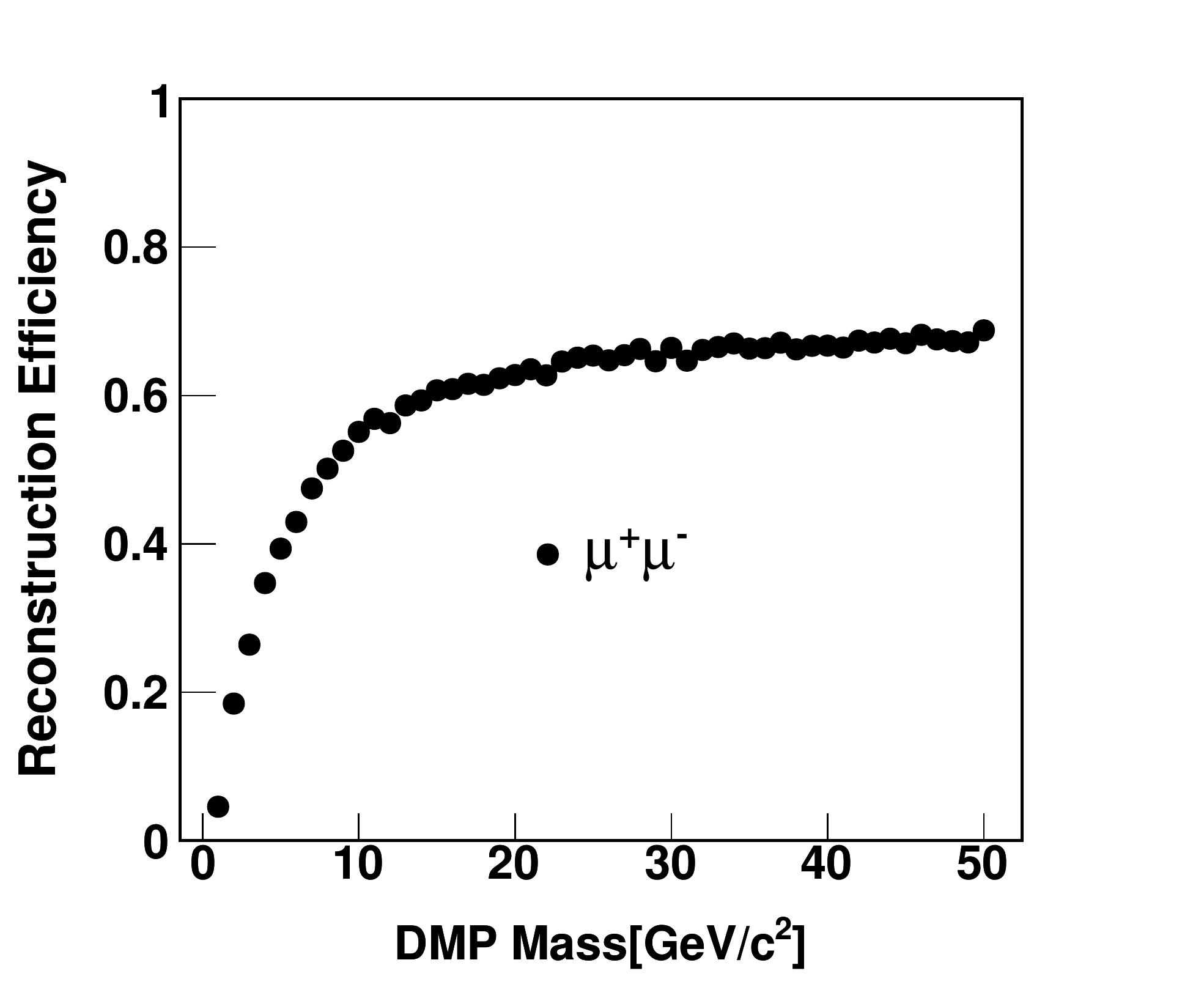}
\includegraphics[width=0.49\textwidth]{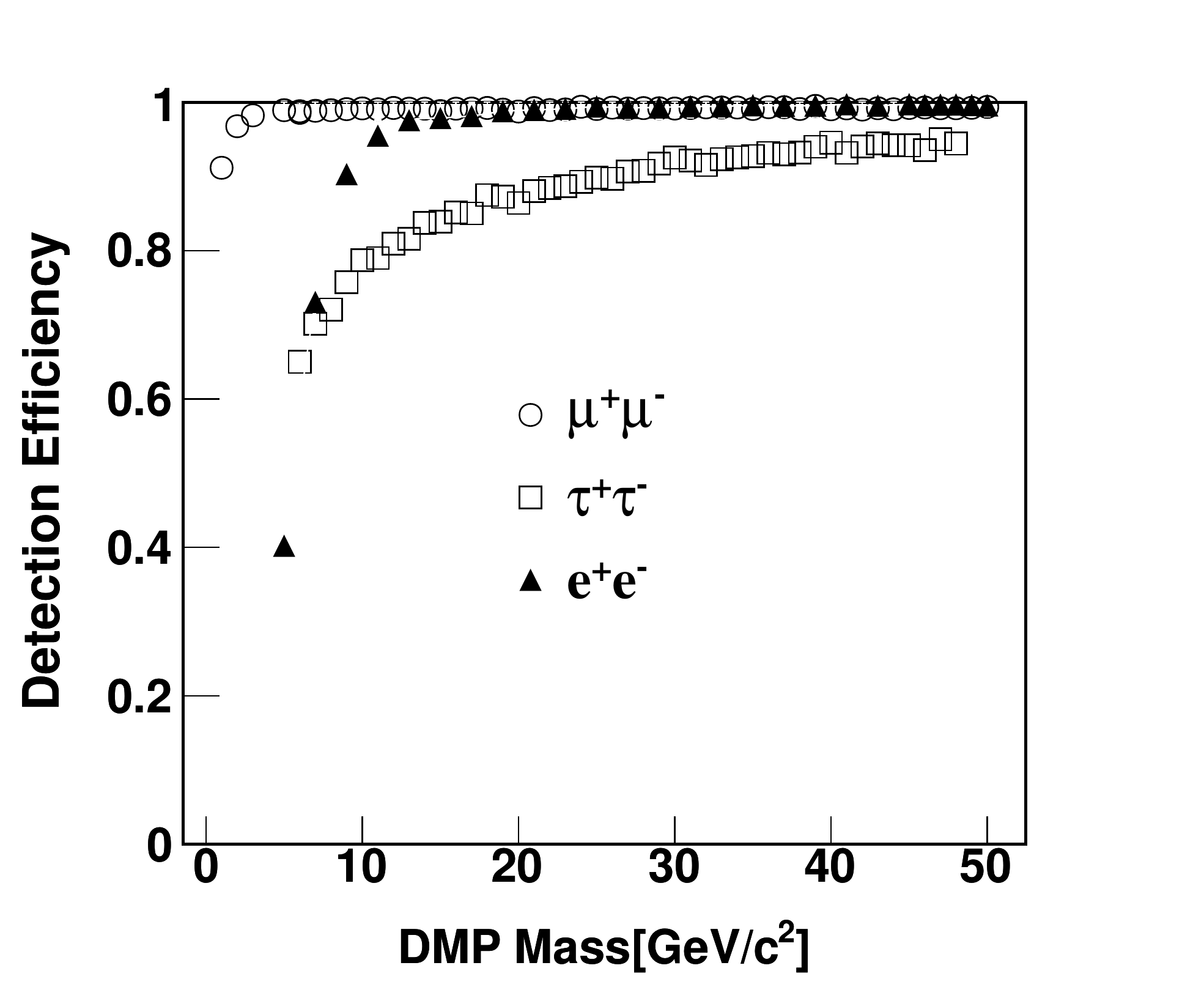}
\caption{Left Panel: The reconstruction efficiency for DMP decays to $\mu^+\mu^-$. Right Panel: The detection efficiency for DMP decays to $\rm{e}$, $\mu$, and $\tau$ channel in air region of the ICAL cavern as marked by different markers inside the plot.}
\label{effinair}
\end{figure}

In case the DMP decays to a pair of muons, one of them will give rise to a clean track in the ICAL detector and the other one will have hits in the side detector in the opposite direction. The background for such events will be the cosmic ray muon or a muon produced due to the interaction of neutrino with the rock and detection in the ICAL detector. But this can be eliminated by using the timing information from the detector. The genuine events are selected by considering the reconstructed momentum within $\pm$3 times the incident momentum measured by the ICAL detector. The Kalman Filter algorithm is used to reconstruct the momentum of the muon inside the ICAL. It is also possible to reconstruct the vertex position and direction cosine inside ICAL. The position of the other muon in the scintillator is obtained by extrapolating the hit position in it using the reconstructed vertex and direction cosine. If the extrapolated position and the simulated position match within a certain range then these events are used in the efficiency calculation. The uncertainty in the measurement of the extrapolated position is taken as $1~\rm{m}$ individually for each position component. The reconstruction efficiency is as shown in the left panel of Fig.{~\ref{effinair}}. With increase in mass the efficiency increases. At lower mass the efficiency is small due to large uncertainty in angular resolution during the extrapolation of the hit position on the side detector. A single layer of the side detector (SD) may not help to distinguish them from background. So to improve this situation at least 2 layers of detector should be in place. From the timing information it will be possible to identify them from others. At least 6-8 layers are needed for good range energy resolution. In this case the efficiency will be obtained by separately reconstructing the direction for SD and ICAL and then by reconstructing the vertex. It will also result in enhancement in the efficiency.

\begin{figure}[h]
\centering
\includegraphics[width=0.49\textwidth]{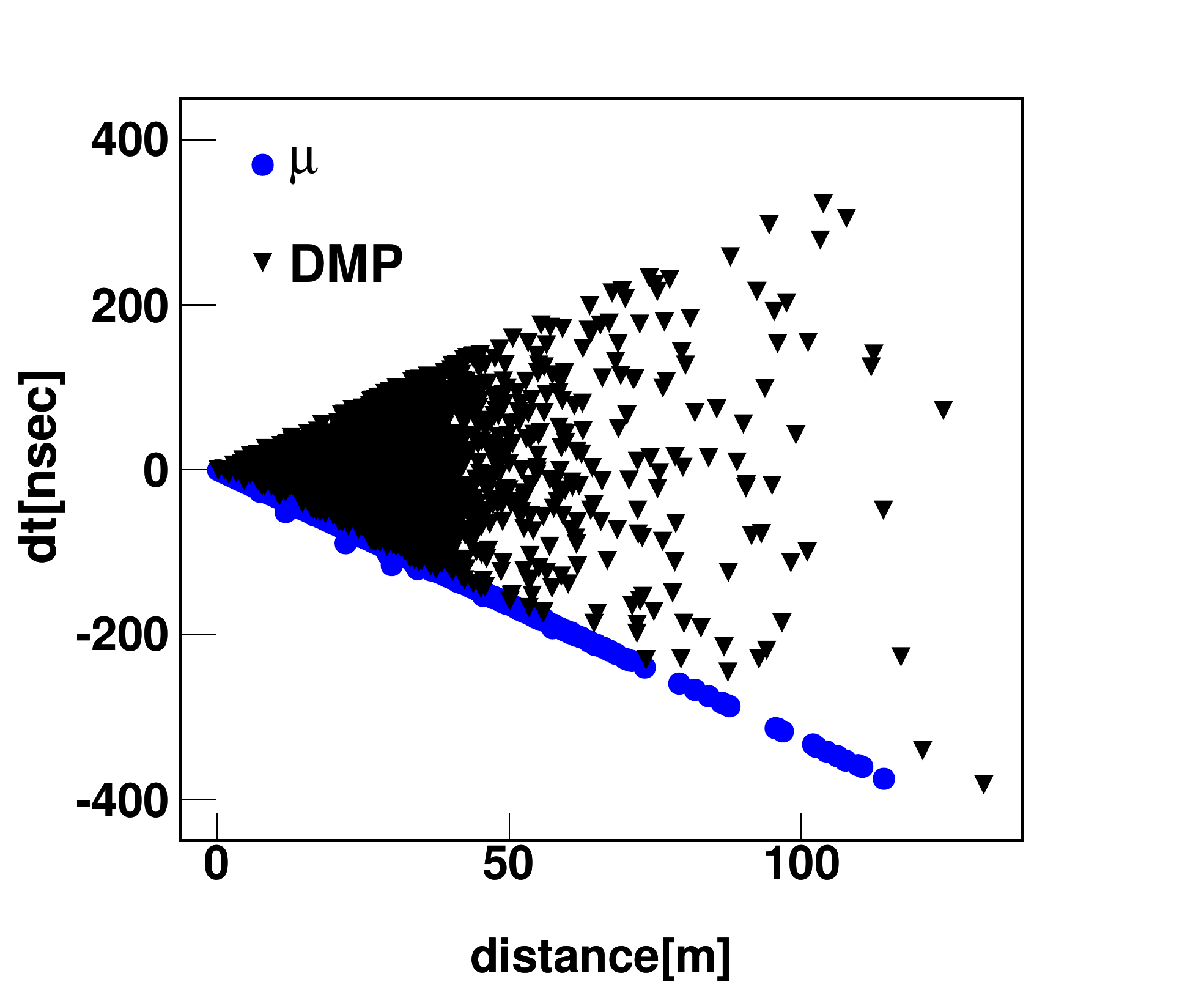}
\caption{The distance between hit points verses time difference.}
\label{tof}
\end{figure}

When DMP decays to a $\tau$ pair, either it will have a bunch of hits or a clean muon track depending on its decay products in the detector. In this case the events produced due to the interaction of the neutrino with the rock matter will be act as a background. So if the DMP vertex is near to the SD, then the background arises from neutrino induced events. This can be identified and eliminated by using time of flight method as shown in Fig.{~\ref{tof}}. It is obtained by measuring the distance between hit coordinates in the two SDs and their corresponding time difference. The cosmic muons originate from outside the ICAL cavern while those that arise from DMP decay are generated inside the ICAL cavern. By excluding a small region near the SD obtained from the simulation for this particular channel we exclude the cosmic muon events. This corresponds to a region close to the edges of the region defined by the 2 delimiting lines in Fig.{~\ref{tof}}. A similar procedure can be applied for the other two channels.

The detection efficiencies are also obtained for DMP decays to $\rm{e}^+\rm{e}^-$ and $\tau^+\tau^-$ mode. They are obtained by using the minimum number of hits in the ICAL detector and is represented in the right panel of the Fig.{~\ref{effinair}}. Even though it is difficult to identify $\rm{e}^\pm$ from hadrons, which mainly lead to showers, the detection probability grows with increasing mass of DMP and saturates beyond $\sim$ $20~\rm{GeV/c^2}$. The detection efficiency for $\mu^\pm$ channel obtained in similar way is also shown in the same plot.

\subsection{Simulation in ICAL}

In this case all the events are generated inside the ICAL detector within a fiducial volume of $40~\rm{m}~\times~14~\rm{m}~\times~12~\rm{m}$. Inside the ICAL it is possible to measure the energy of two muons and hence the invariant mass.

\begin{figure}[h]
\centering
 {\includegraphics[width=0.49\textwidth]{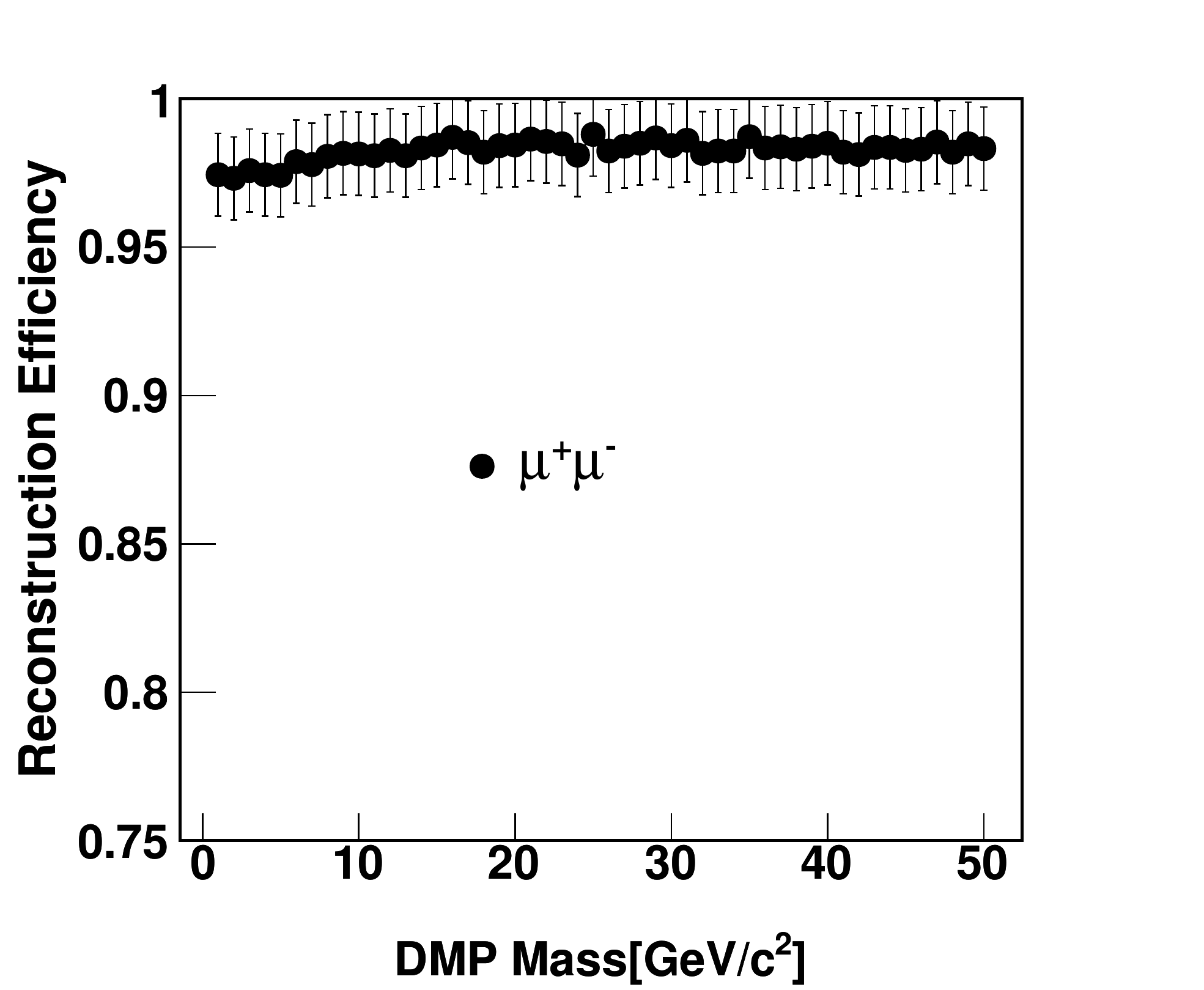}}
\caption{The detection efficiency inside the ICAL for the 
DMP decays to $\mu^+\mu^-$ pair.}
\label{effinical}
\end{figure}

For DMP decays to $\mu^+\mu^-$ the timing and trajectory information can be used to separate them from background like cosmic ray muon and neutrino. The Monte-Carlo technique is used to simulate DMP decays, uniformly distributed with the fiducial volume of ICAL, and track the decay muons. For each energy and theta bin the momentum resolution and direction resolution are used separately for $\mu^+$ and $\mu^-$ from muon look-up table. Figure{~\ref{effinical}} shows the detection efficiency for DMP decays to $\mu^+\mu^-$ channel inside the ICAL.

The GEANT4 simulation tool-kit is used to do the simulation for DMP decays to $\tau^+\tau^-$ channel. As $\tau$ decays to other particles in two- and three-body decay modes on a time scale of $\sim$ psec, it is easier to do the simulation in GEANT4 instead of a simple Monte-Carlo simulation. Inside the ICAL the background for such events will be arise from neutral current (NC) interactions induced by neutrinos. However the main difference between them is that in a NC event all the particles will be in one direction, whereas in DMP decay the decay products will be on either side of a single vertex resulting in an increase in the time with hit position in both the directions.

\subsection{Detector acceptance} 

\begin{figure}[h]
\centering
 {\includegraphics[width=0.49\textwidth]{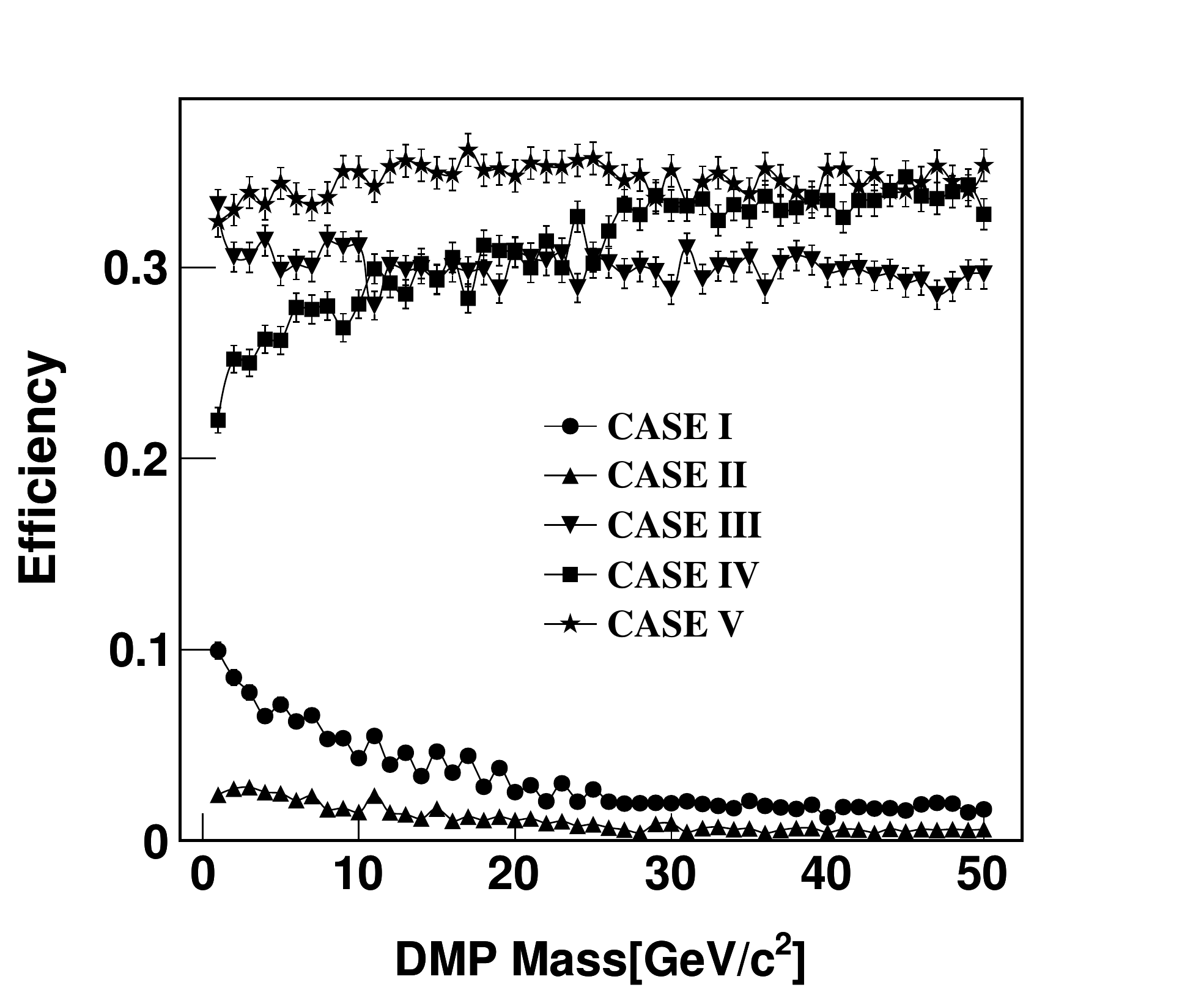}}
\caption{The detector acceptance by considering the DMP decays to $\mu$ pair in the whole ICAL cavern. {\textsc{case} I}: Not detected, {\textsc{case} II}: 2 or 1 in ICAL not in SD, {\textsc{case} III}: 2 in SD and not in ICAL, {\textsc{case} IV}: 1 in ICAL and 2nd one in SD, and {\textsc{case} V}: 1 in SD and another one is not detected.}
\label{acceptance}
\end{figure}

In the above two cases, we forced the particles to be generated in their respective regions. But to get the detector acceptance, the events are generated uniformly whole over the ICAL cavern i.e. including the ICAL detector and the air region. Only the DMP decays to $\mu^\pm$ channel are used and the simulation is carried out using GEANT4. Figure {~\ref{acceptance}} shows the detection efficiency for 5 different situations. Bands with different markers represent 5 different cases. The classification is based on their type of detection. The efficiency is obtained by taking the ratio between the numbers of events with hit in the respective detector to the total number of simulated events.

The cases II and IV are sensitive to DMP decay detection. The Case III is also sensitive if absorbers are placed between SD and is relevant for low mass of DMP.


\section{Analysis and Results}

\begin{figure}[h]
\centering
\includegraphics[width=0.49\textwidth]{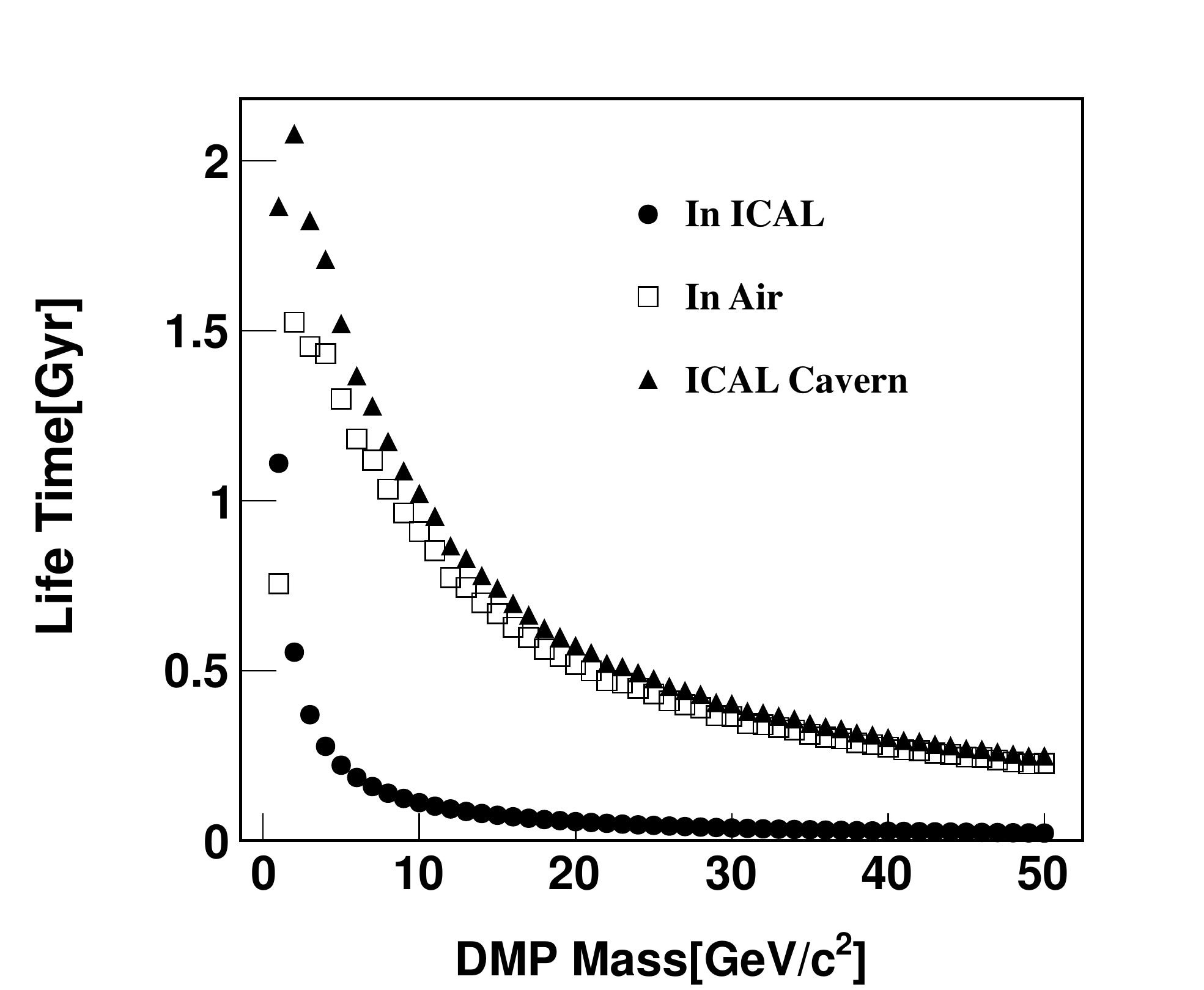}
\includegraphics[width=0.49\textwidth]{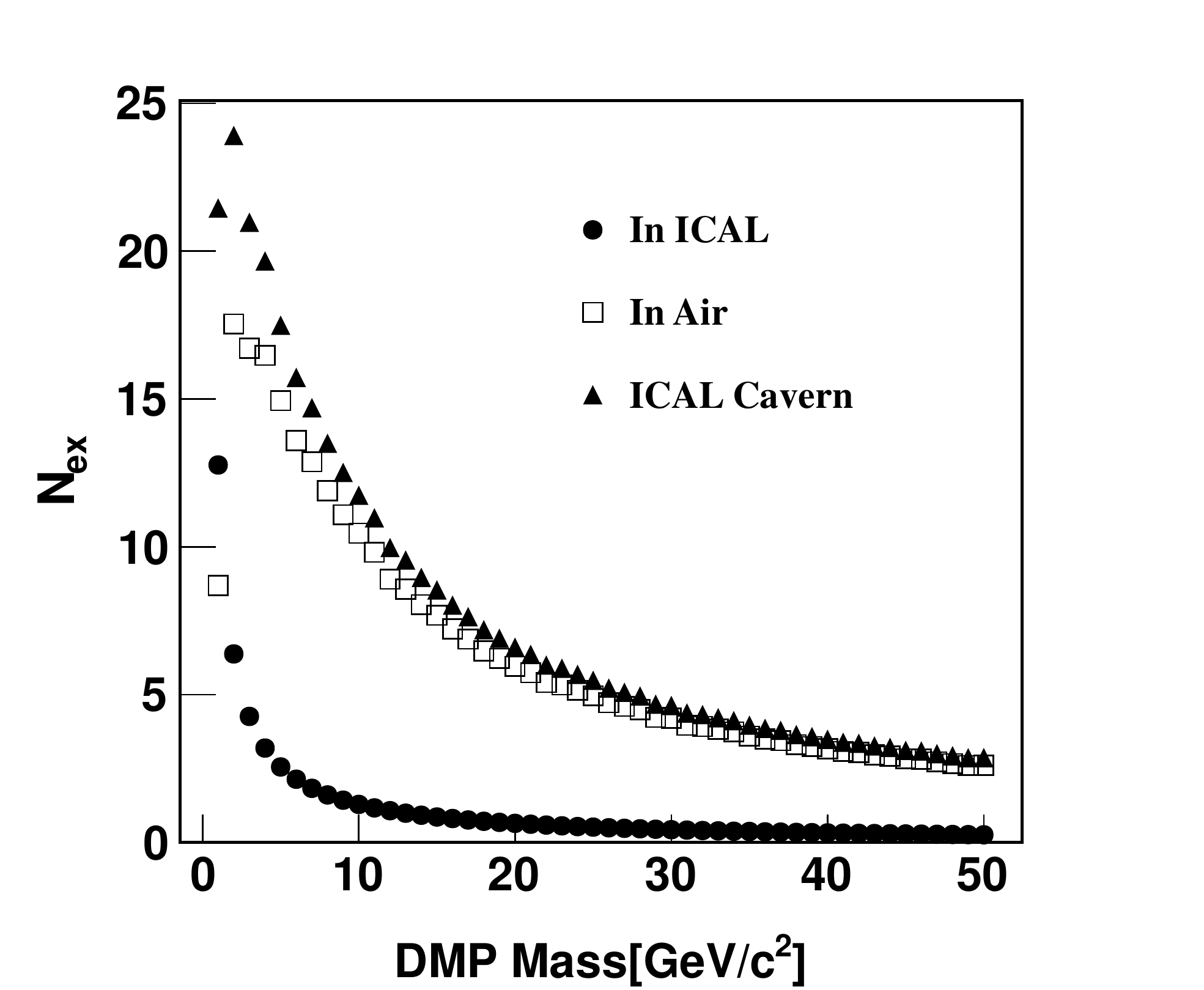}
\caption{Left Panel: The lower limit in the life time of DMP verses its mass for $\mu^+\mu^-$ decay channel. Right Panel: The number of expected events due to the DMP decays to $\mu^{\pm}$ with life time $2~\rm{Gyr}$ at ICAL cavern for $10~\rm{yrs}$ of detector running period.}
\label{lifetime_event}
\end{figure}

The obtained efficiency in different regions from Sec.4 is used to estimate the life time of a DMP for a finite number of observed events using Frequentist Method \cite{pdg}. If $\rho$ (GeV/$\rm{cm}^3$) is the local dark matter density, $\rm{V}$ ($\rm{cm}^3$) is the detection volume, $\epsilon$ is the detection efficiency, $\rm{B}$ is the branching ratio, $\rm{M}$ (GeV/$\rm{c}^2$) is the DMP mass and $\rm{R}~(\rm{Yr}^{-1})$ is the rate of decay, then the life time T ($\rm{Yr}$) of the DMP is given by
\begin{equation} 
\label{lifetime}
\rm{T} = \frac{\rho~\rm{V}~\epsilon~\rm{B}}{\rm{M}~\rm{R}}.
\end{equation}
Here and for further calculation the local dark matter density is taken as $0.39~\rm{{GeV}/{cm^3}}$ \cite{pdg}. The limit on the life time is obtained by considering the upper limit for the 0 observed events with $90\%$ confidence level for 0 background in 1 year of detector running time using Eq.(\ref{lifetime}). The obtained limit for different regions with mass is shown in the left panel of Fig.{~\ref{lifetime_event}} by different symbols. With increase in the mass of DMP, the limit on the life time of the DMP is decreasing due to the decrease in number density in a fixed volume. From Super-Kamiokande (SK) \cite{superk} results there is a stringent limit on the life time of the dark matter particle decaying to $\mu^{\pm}$ pair obtained from $\nu_{\mu}$ signal starting with mass 10~$\rm{GeV/c^2}$. For 10~$\rm{GeV/c^2}$ DMP mass the life time is of the order of $10^7$~Gyr as obtained by SK. The ICAL can also put bound on the DMP life time in the similar way as followed by the cherenkov detectors and will be comparable to them. 

Alternatively, using the higher limit of DMP life time (2 Gyr) from the above result and the local dark matter density, the number of expected events due to the decay of DMP is obtained separately for events simulated in air region between cavern wall and ICAL detector, inside the ICAL detector and the whole of the ICAL cavern by combining the first two results. The expression for expected number of events ($\rm{N_{ex}}$) due to the decay of DMP is given by
\begin{equation}
\label{events}
\rm{N_{ex}}= \frac{\rho~\rm{V}~\epsilon~\rm{B}}{\rm{M}~\rm{T}}.
\end{equation}
The right panel of the Fig.{~\ref{lifetime_event}} shows the number of expected events for 10 years of counting for the 3 cases mentioned above.

\section{Conclusions}

If the reinterpretation of the Kolar events as being due to the decay of dark matter particles is correct, it should be possible to observe them with much larger numbers in the ICAL cavern at the India-based Neutrino Observatory. Due to the larger volume of the ICAL cavern and the size of the ICAL detector, the expected number of such events is doubled compared to those observed at Kolar Gold Fields. In addition to addressing the issues towards understanding of the anomalous Kolar events, a non-observation of such events at INO will be able to place bounds on the life time of the dark matter particles with mass between $1~\rm{GeV/c^2}~\rm{and}~50~\rm{GeV/c^2}$. In another way it is estimating limit on the life time of the DMP with mass below 10~$\rm{GeV/c^2}$, however Super-Kamiokande has put limit from 10~$\rm{GeV/c^2}$ to few TeV.


\section*{Acknowledgements}
We would like to thank Prof. M.V.N. Murthy for his critical comments and suggestions. We also wish to thank Prof. D. Indumathi and Prof. G.Rajasekaran for their interest in the work reported in this paper. We are also grateful to Prof. Pijushpani Bhattacharjee and Prof. Nita Sinha for their valuable comments and suggestion during the preparation of the manuscript. Our thanks are also to the INO collaborators for their invaluable support. 


\end{document}